\documentclass[prb,onecolumn,showpacs,showkeys,amsmath,amssymb]{revtex4-1}
\usepackage{refstyle}
\usepackage{graphicx}
\usepackage{dcolumn}
\usepackage{tabularx}
\usepackage{bm}
\usepackage[section]{placeins}
\usepackage{subfigure}

\begin{document}

\title{Ferromagnetic ordering of minority Ce$^{3+}$ spins in a quasi-skutterudite Ce${_3}$Os${_4}$Ge$_{13}$ single crystal}
\author{Om Prakash}
\email{op1111shukla@gmail.com}
\affiliation{Department of Condensed Matter Physics and Materials Science,
Tata Institute of Fundamental Research, Mumbai-400005, India}
\author{A. Thamizhavel}
\affiliation{Department of Condensed Matter Physics and Materials Science,
Tata Institute of Fundamental Research, Mumbai-400005, India}
\author{S. Ramakrishnan}
\affiliation{Department of Condensed Matter Physics and Materials Science,
Tata Institute of Fundamental Research, Mumbai-400005, India}

\begin{abstract}
We report site disorder driven ferromagnetic ordering of nearly 8$\%$ Ce atoms in single crystalline Ce${_3}$Os${_4}$Ge$_{13}$ below 0.5~K. Ce${_3}$Os${_4}$Ge$_{13}$ crystallizes in a cubic structure with space group $\it{Pm\bar{3}n}$. The structural analysis shows the presence of site-disorder in the system, where 2(a) Ge site is partially occupied by the Ce atoms. Due to the small Ce $4f$-ligand hybridization, these Ce atoms are in the Ce$^{3+}$ 
(magnetic) state while rest of the Ce atoms in the unit cell are in Ce$^{4-\delta}$ (non-magnetic) state. The heat capacity shows a peak below 0.5~K corresponding to the ferromagnetic ordering of the Ce$^{3+}$ moments. The ferromagnetic ordering below 0.5~K is also seen in the ac-susceptibility data. At low temperatures (1.8~K$\leq T\leq 6$~K), the magnetization shows $log(T)$ dependence, whereas the resistivity shows power law ($T^n$) temperature dependence indicating the non-Fermi liquid behavior of the quasiparticles. 
\end{abstract}
\maketitle
\section{Introduction}
Cerium based compounds show interesting ground state properties such as local moment and band magnetism, unconventional superconductivity, Kondo-effect, coherent Fermi-liquid and non-Fermi liquid behavior, and quantum criticality, owing to the interplay of valence fluctuation, heavy fermion and crystal field splitting \cite{Zhou2013,Petrovic2001,Felner1997,Malik1996,STEGLICH1991,Lohneysen1994,Cornut1972}. Variation in the cerium valence state plays a crucial role in the ground state properties of Ce-compounds. Among the Ce-based compounds studied till date, most of the compounds have single crystallographic site for Ce-ion within the unit cell. The compounds with multiple inequivalent crystallographic Ce sites in a unit cell show complex electronics and magnetic phenomena \cite{Prokleska2015,Feyerherm2012,Fikek2012,Fikek2013,Proke2014,Ghosh1995,BOUCHERLE1989,Bonnet1994,Lawrence1991,FUKUHARA1992,Godart1993,Godart1994,Feyerherm2012}, as Ce-ions in inequivalent sites are subjected to different local environment resulting in different interactions of Ce $4f$-electrons and ligands. Recently, 
the work of Prokle\ifmmode \check{s}\else \v{s}\fi{}ka et. al. \cite{Prokleska2015} shows that the presence of Ce at multiple sites leads to two antiferromagnetic transitions and superconductivity in a heavy fermion compound ${\mathrm{Ce}}_{3}{\mathrm{PtIn}}_{11}$. Recent studies on the single-crystals as well as poly-crystals of CeRuSn suggest 50\% of the Ce atoms undergo antiferromagnetic ordering while the rest remain the Ce$^{4-\delta}$ state \cite{Feyerherm2012,Fikek2012,Fikek2013,Proke2014}. Similarly, polycrystalline Ce${_3}$Ru${_4}$Ge$_{13}$, having two inequivalent Ce sites (Ce$^{3+}$ and Ce$^{4-\delta}$ states at different sites) was shown to exhibit spin glass behavior \cite{Ghosh1995}.

The compounds with Ce$^{3+}$ magnetic moments have two competing interactions, {\it{viz.}} Ruderman-Kittel-Kasuya-Yosida (RKKY) exchange interaction mediated by conduction electrons \cite{Ruderman1954,Kasuya1956} and on-site Kondo interaction \cite{Kondo1964}. At low temperatures, strong RKKY exchange interactions yields a magnetic ground state while strong Kondo interactions screen the Ce$^{3+}$ moments (conduction electrons form singlet state with the Ce$^{3+}$ moments) leading to a non-magnetic ground state \cite{DONIACH1977}. The rare-earth quasi-skutterudites R$_3$T$_4$X$_{13}$ compounds, where R is rare earth, T is transition metal from 3d to 5d, and X is an III-IV group element, provide an ideal playground to study these phenomena \cite{Yang2015,om2015,Sato1993}.
	
Here, we present low temperatures studies on the physical properties of a cubic quasi-skutterudites compound Ce$_3$Os$_4$Ge$_{13}$ (space group \# 223). Ce$_3$Os$_4$Ge$_{13}$ has two inequivalent Ce-sites (2(a) and 6(d), where 2(a) has a cubic site symmetry) driven by site-disorder. We find that approximately 8\% of the total Ce ions are present on the 2(a) site. While the Ce ions on the 2(a) site are in Ce$^{3+}$ (magnetic) state, those on the 6(d) site are in the Ce$^{4-\delta}$ (non-magnetic) state. The ac-susceptibility and heat capacity measurements show that these Ce$^{3+}$ moments at 2(a)-site order ferromagnetically below 0.5~K. The low-temperature magnetization data above 1.8~K shows $log(T)$ behavior, whereas the resistivity data shows power-law temperature dependence, suggesting the non-Fermi liquid type behavior in Ce${_3}$Os${_4}$Ge$_{13}$. 
\section{Experimental Details}
Ce${_3}$Os${_4}$Ge$_{13}$ single crystal was grown using Czochralski crystal pulling method in a tetra-arc furnace under inert argon atmosphere. The stoichiometric mixture (10~g) of highly pure elements (Ce: 99.99\%, Os: 99.99\%, Ge: 99.99\%) was melted 3-4 times in the tetra-arc furnace to make a homogeneous polycrystalline sample. The single crystal was pulled using a tungsten seed rod at the rate of 10~mm/h for about 6~hrs to get 5-6~cm long and 3-4~mm diameter cylindrical shaped crystal. The phase purity of the crystal was characterized by room temperature powder x-ray diffraction (PXRD) using PAN-alytical x-ray diffractometer utilizing monochromatic Cu-$K_{\alpha}$ radiation ($\lambda= 1.5406\AA$). The Electron Probe Micro-Analysis (EPMA) and Energy Dispersive x-ray Spectroscopy (EDX) characterizations were done on the polished surfaces and confirm proper stoichiometry [3-4-13] and single phase nature of Ce${_3}$Os${_4}$Ge$_{13}$ compound. The single crystal was oriented along the crystallographic direction [100] using Laue back reflection method in a Huber Laue diffractometer and cut to the desired shape and dimensions using a spark erosion cutting machine. The electrical resistivity was measured using the standard four-probe technique in Physical Property Measurement System (PPMS, Quantum design, USA) from 0.05-300~K in various magnetic fields. Gold (Au) wires were attached to the sample using silver paste for making ohmic contacts. The magnetic susceptibility was measured in commercial SQUID magnetometer (MPMS5, Quantum Design, USA) from 1.8-300~K in different magnetic fields under zero field cooled (ZFC) and field cooled (FC) conditions. The isothermal magnetization was measured from $-14$~T to $+14$~T at 2-8~K temperatures with 1~K temperature interval. The single crystal oriented along [100] crystallographic direction was used for this measurements with the magnetic field (H) being parallel to [100]. The heat capacity ($C_p$) was measured by the thermal relaxation method using PPMS from 0.05 to 150~K in 0-5~T magnetic fields. For temperatures below 2~K, we used dilution insert of the PPMS for the heat capacity as well as resistivity measurements. The ac-susceptibility measurement was done in a CMR-mili-kelvin setup using a gradiometer pickup coil from 0.1-4.2~K.

\begin{table}[htp]
\caption{\small{ Crystal structure parameters obtained from the Rietveld refinement of the room 
temperature powder x-ray diffraction data of the unit cell of cubic Ce$_{3}$Os$_{4}$Ge$_{13}$. Profile reliability factor $R_{p} = 17.5\% $, 
weighted profile $R$-factor $R_{wp}=17.2\%$, Bragg R-factor $= 8.46\%$ and R$_{f}$-factor $= 7.44\%$ were obtained from the best fit.
}}
\begin{ruledtabular}
\begin{tabularx}{0.35\textwidth}{cccccc}
\multicolumn{2}{l}{Structure} &\multicolumn{3}{l} {Cubic} \\
\multicolumn{2}{l}{Space group} & \multicolumn{2}{l} {$\it{Pm\bar{3}n}$ (No. 223)}\\
\multicolumn{2}{l}{Lattice  parameters} \\
\hline
\multicolumn{2}{l}{\hspace{0.8cm} $a$ ({\AA})}&  8.905(1)  \\        
\multicolumn{2}{l}{\hspace{0.8cm}$V_{\rm cell}$({\AA}$^{3}$)} &706.16( 0.03)  \\
\\
Atom&Wyckoff posi.&$x$&$y$&$z$&Occ.\\
\hline\hline
Ce 1 &6(d)&0.25000&0.50000&0.00000&5.500 \\

Ce 2 &2(a)&0.00000&0.00000&0.00000&0.500\\

Os &8(e)&0.25000&0.25000&0.25000&8.000 \\

Ge 1 &24(k)&0.00000&0.68487	&0.84788&24.00\\

Ge 2 &2(a)&0.00000&0.00000&0.00000&1.453\\

Ge 3&6(d)&0.25000&0.50000&0.00000&0.547\\        
\hline
\end{tabularx}
\end{ruledtabular}
\label{table:1}
\end{table}

\section{Results and Discussion}
\begin{figure}[htbp]
\includegraphics[width=7cm,angle=0]{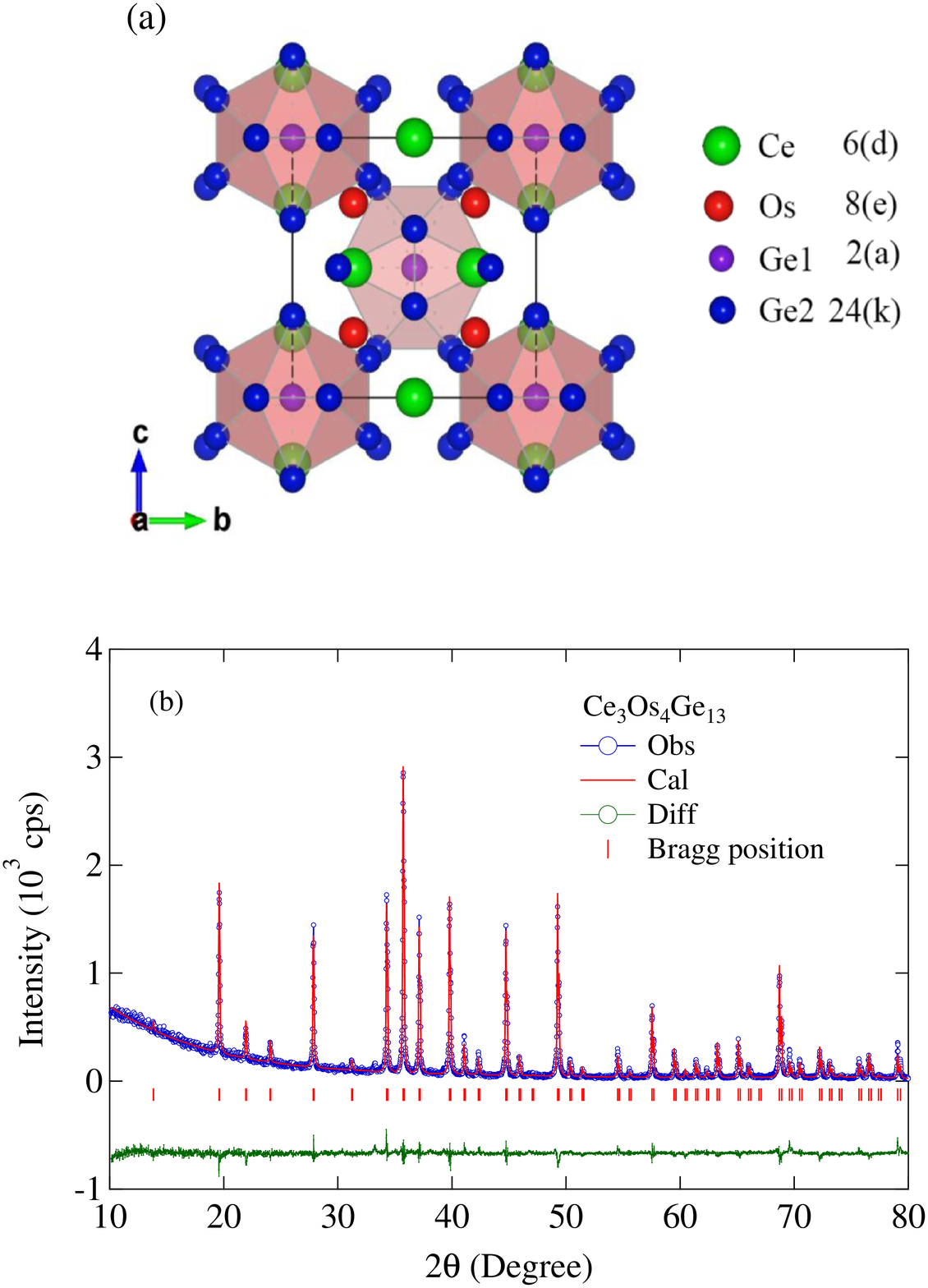}
\caption{(Color online) (a) Crystal structure of a unit cell of Ce${_3}$Os${_4}$Ge$_{13}$ projected along (100) plane. The unit cell consists of cage-like sub-structures which contain Ge 2(a) sites inside. (b) Rietveld refinement of the powder XRD data of Ce${_3}$Os${_4}$Ge$_{13}$. The Rietveld analysis confirms the single phase nature of the grown crystal.}
\label{fig:fig1}
\end{figure}

Figure~\ref{fig:fig1} shows the PXRD analysis and crystal structure of Ce${_3}$Os${_4}$Ge$_{13}$. The crystal structure of Ce${_3}$Os${_4}$Ge$_{13}$ was reported to be cubic (space group $\it{Pm\bar{3}n}$) by Braun and Segre \cite{Braun1981} and it is closely related to the R${_3}$Rh${_4}$Sn$_{13}$ structure reported by Remeika et al  \cite{Remeika1980}. Our x-ray diffraction measurements on the powder sample confirm the crystal structure and the space group. The unit cell consists of cage-like substructures and contains 40 atoms (2 formula unit). The unit cell has two inequivalent Ge sites namely 2(a) $\&$ 24(k). The analysis of the room temperature powder XRD data using structural Rietveld refinement \cite{Carvajal1993} (using the Fullprof program) confirmed the single phase nature of the Ce${_3}$Os${_4}$Ge$_{13}$ single crystal. There is a small peak around 33 degrees in the x-ray diffraction pattern which is not fitted with the Rietveld model for the space group $\it{Pm\bar{3}n}$ (No. 223). The exact origin of this peak is not clear at the moment but it can be due to the splitting of the Ge2 24(k) site or disorder at the Ge1 2(a) site as found in many isostructural compounds \cite{Gumeniuk2012,Gumeniuk2015}. The Rietveld analysis shows the existence of site disorder in the structure with 0.25 Ce atoms occupying the 2(a) (Ge) sites per formula unit. Similar site disorder had been reported in other isostructural compounds \cite{Ghosh1995}. The lattice parameters obtained from Rietveld refinement are listed in Table \ref{table:1}.

Cerium can exist in Ce$^{3+}$ (magnetic) and/or Ce$^{4-\delta}$(non-magnetic/mixed-valence) valence states depending on the hybridization strength of the Ce $4f$-electron and valence states of the ligands \cite{Fujimori1988,Tien2000,LIANG2002}.  The hybridization strength depends on the overlap of the Ce $4f$-electron wave function and the ligand valence electron wave functions \cite{LIANG2002}. If the distance between the Ce ion and ligand is sufficiently large, the magnetic Ce$^{3+}$ state is preserved due to small $4f$-ligand hybridization. Conversely, by decreasing the distance between Ce ion and ligand, the $4f$-ligand hybridization can be tuned such that Ce$^{3+}$ ion loses its only $4f$-electron and becomes  mixed-valent, Ce$^{4-\delta}$. The co-existence of Ce$^{3+}$ and Ce$^{4-\delta}$ states is known as the intermediate valence state. The Ce intermediate valence ions do not carry stable magnetic moment \cite{Malterre1989} and hence do not show magnetic ordering. The Ce at 2(a) site is loosely bound within the cage, hence is weakly hybridized with other atoms as compared to the Ce at 6(d) site. This leads to the preservation of Ce$^{3+}$ ions at these 0.25 occupancies at 2(a) site per formula unit of Ce${_3}$Os${_4}$Ge$_{13}$. The 6(d) site favours the presence of Ce$^{4-\delta}$ ions with remaining 2.75 occupancy per formula unit due to stronger hybridization. As Ce$^{3+}$ ions at 2(a) sites are within cages and well separated from other each other, the strength of the magnetic interactions is expected to be very weak. 

\begin{figure}[h]
\includegraphics[width=7cm,angle=0]{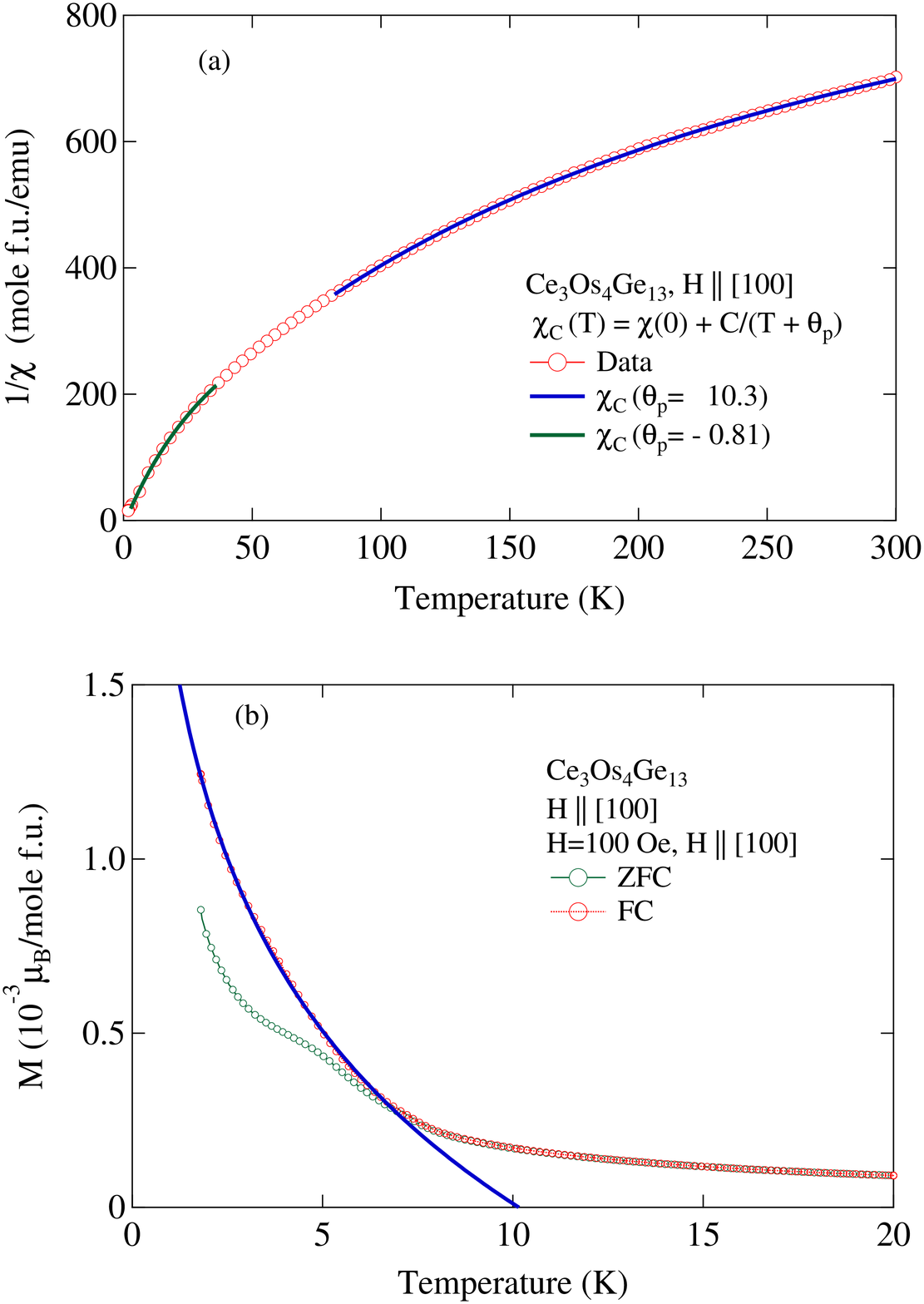}
\caption{(Color online) (a) Temperature dependence of inverse dc-susceptibility of Ce${_3}$Os${_4}$Ge$_{13}$ for H $\parallel [100]$ from 1.8-300~K. The data has been fitted to the Curie-Weiss equation at low temperature $1.8<T<30$~K and high temperature $100<T<300$~K separately. The value of $\theta_{p}$ changes from 10.3~K to -0.81~K going from high temperature to low temperatures suggesting that the magnetic correlations are antiferromagnetic in nature at higher temperatures and ferromagnetic at low temperatures. (b) Magnetization as a function of temperature in the FC and ZFC condition with $H=100$~Oe. The continuous curve is the logarithmic fit $M(T)=a-b\log(T)$ to the FC data. }
\label{fig:fig2}
\end{figure}

Now, the question arises, ``what is the ground state of these minority Ce$^{3+}$ moments at low temperatures?". To understand the interaction between the Ce$^{3+}$ moments as a function of temperature, we have measured dc-susceptibility of Ce${_3}$Os${_4}$Ge$_{13}$ from 1.8-300~K. The ac-susceptibility as a function of temperature is measured from 0.1-4.2~K to study nature of magnetic correlations in Ce${_3}$Os${_4}$Ge$_{13}$ at low temperatures. The temperature dependence of the inverse dc-susceptibility from 1.8~K to 300 K is shown in Fig.~\ref{fig:fig2}(a). A rapid rise in the dc-susceptibility below 20~K is observed. To estimate the value of the effective magnetic moment in the compound, the high-temperature susceptibility data from 85 to 300~K is fitted to the Curie-Weiss expression (see Fig.~\ref{fig:fig2}(a)),
\begin{equation}
\label{eqn1}
\chi = \chi (0) + C/(T + \theta_{p})
\end{equation}
where $\chi(0)$ represents the temperature-independent part of the magnetic susceptibility, including the core-electron diamagnetism and the Pauli paramagnetism and VanVleck terms \cite{VanVleck1932}, C is the Curie constant and $\theta_{p}$ is the paramagnetic Curie-Weiss temperature. From the best fit, we get $\chi(0) = 8.52\pm{0.01} \times 10^{-4}$~emu/mol, C$=0.18$ emu-K/mole-f.u., $\theta_{p} = 10.3\pm{0.1}$~K and $\mu_\mathrm{eff}= 0.69\pm{0.02} \mu_\mathrm{B}$/mole-f.u.. If we assume that each Ce$^{3+}$ ion has a Curie constant of 0.807 emu-K, the expected value of the Curie constant is 0.807 emu-K$\times$0.25/mol-f.u. = 0.201 emu-K/mole-f.u. (since we have only 0.25 Ce$^{3+}$ moments per f.u. in Ce${_3}$Os${_4}$Ge$_{13}$), which is close to the observed value of the Curie constant C$=0.18$ emu-K/mole-f.u.. This value of $\mu_\mathrm{eff}$ is much smaller the value of 2.54$\mu_\mathrm{B}$ expected for the free ion moment of Ce$^{3+}$. In a perfectly ordered crystal with the occupation of 6 Ce atoms at the 6(d) sites in the unit cell, the susceptibility should be temperature independent reflecting the presence of only Ce$^{4-\delta}$ ions. The existence of a non-zero value of the effective moment of Ce clearly shows the structure is not perfectly ordered (presence of site disorder), which is in agreement with the Rietveld analysis of the powder x-ray diffraction data, as discussed above. The Rietveld refinement suggests that the occupancy at 2(a) site in the unit cell is shared between 0.5 Ce atoms and 1.5 Ge atoms (leading to 0.25 Ce occupancy on Ge 2(a) site per formula unit). These 0.25 Ce atoms per formula unit at 2(a) site are likely to be in Ce$^{3+}$ state because of larger Ce-Ge distances in the cage like substructure. If we assume that only 0.25 Ce atoms per formula unit at the 2(a) site contribute to the magnetization, then the effective moment can be written as, $\mu_\mathrm{eff} = 2.54*0.25$, resulting in a effective magnetic moment value 0.64$\mu_\mathrm{B}$, which is very close to the experimentally observed value of 0.69$\mu_\mathrm{B}$ above 100~K. This agreement in the effective magnetic moment values along with the Rietveld analysis results go hand in hand to suggest that the Ce atoms which are present at the 2(a) sites in the unit cell are magnetic (Ce$^{3+}$) and rest of the Ce atoms at 6(d) are non-magnetic (Ce$^{4-\delta}$). The positive value of $\theta_{p} = 10.3$~K suggests presence of antiferromagnetic correlations at higher temperatures in this compound. The Curie-Weiss fit to the inverse dc-susceptibility below 30~K gives negative $\theta_{p} = -0.81$~K indicating crossover in the nature of magnetic correlations from antiferromagnetic to ferromagnetic as the sample is cooled to lower temperatures. The value of the Curie constant below 30~K is C$=0.1$ emu-K/mole-f.u..

This change in the nature of magnetic correlations can be understood as following: At higher temperatures we have Kondo-screening of Ce$^{3+}$ magnetic moments by conduction electrons. This coupling of conduction electrons with magnetic moments is known to be antiferromagnetic in nature. As we cool down the system to lower temperatures the effective screening of the magnetic moments increases. In general, the interaction of conduction electrons and magnetic moment can be classified in three different cases \cite{Stewart2001} in terms of number ($n$) of orbital degree of freedom, or channels of conduction electrons and magnetic impurity spin S. (a) $n=2S$, the number of channels (bands) of conduction electrons is just sufficient to screen the magnetic impurity spin forming a singlet. This is normal Kondo problem giving rise to Fermi-liquid behavior. In this case, the conduction electrons fully screen the local magnetic spin below Kondo temperature $T_K$. (b) $n<2S$, the magnetic spin is not fully compensated since there are not enough conduction electron channels. (c) $n>2S$, the magnetic impurity is overcompensated and a critical behavior sets in as temperature and magnetic field $\rightarrow 0$. This results in the divergence of the length $\xi$, over which the magnetic spin affects the conduction electrons. In this case, power law or logarithmic temperature dependence is observed in measured quantities like magnetization, resistivity or specific heat and a non-Fermi liquid behavior is expected at low temperatures. Furthermore, in the overcompensated case, the local coupling between the Ce$^{3+}$ spins and conduction electrons can give rise to magnetic order via RKKY interaction. In particular, presence of Ce$^{3+}$ moment in a cubic symmetry, as is the case in Ce${_3}$Os${_4}$Ge$_{13}$ (site 2(a) occupying Ce$^{3+}$ moments has cubic site symmetry within the unit cell), is predicted to have a magnetic ground state \cite{COX1993}. At low temperatures, the competition between RRKY interaction and Kondo-screening plays a crucial role in the magnetic nature of the ground state.

The field cooled (FC) and zero field cooled (ZFC) magnetization data in a magnetic field of 100~Oe for Ce${_3}$Os${_4}$Ge$_{13}$ single crystal is presented in Fig.~\ref{fig:fig2}(b). The magnetization in the FC state is larger than ZFC state, clearly showing the presence of ferromagnetic correlations below 5~K. The magnetization increases with decreasing temperature but does not saturate till 1.8~K. This suggests that the ferromagnetic interaction strength is weak in Ce${_3}$Os${_4}$Ge$_{13}$ for $T>1.8$~K, which is expected as the screened Ce$^{3+}$ moments (being inside a cage substructure) are well separated from each other. No signature of magnetic ordering above 1.8~K is observed in transport and heat capacity measurements, suggesting the absence of long-range ferromagnetic ordering in Ce${_3}$Os${_4}$Ge$_{13}$. The FC magnetization data below 5~K is fitted by $M(T)=a-b\log(T)$, where $a=1.65\times10^{-3}$ and $b=1.66\times10^{-3}$, indicating non-fermi liquid behavior in Ce${_3}$Os${_4}$Ge$_{13}$ at low temperatures.

\begin{figure}[h]
\includegraphics[width=7cm,angle=0]{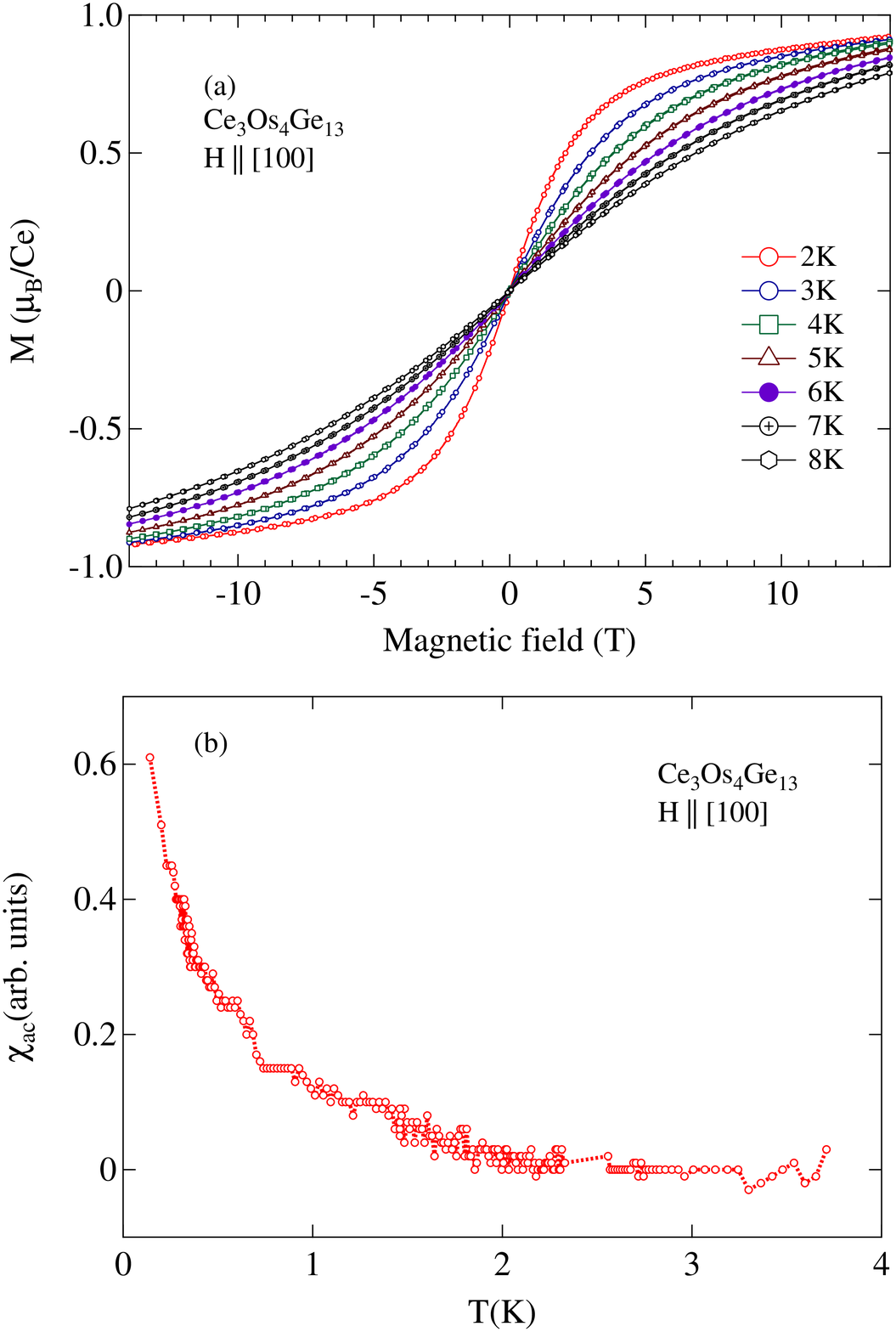}
\caption{(Color online) (a) Iso-thermal magnetization $M(H)$ data at temperatures from 2-8~K for $H\parallel [100]$ direction. (b) Temperature dependence of the ac-susceptibility data shows upturn below 0.5~K indicating ferromagnetic ordering of minority Ce$^{3+}$ moments.}
\label{fig:fig3}
\end{figure}

The isothermal magnetization data measured at different temperatures is shown in the Fig.~\ref{fig:fig3}(a). The saturation of magnetization at high magnetic fields signifies the presence of ferromagnetic correlations in this compound. The ordered magnetic moment of a free Ce$^{3+}$ ion is $g_J J = 2.16 \mu_\mathrm{B}$. The saturation magnetization in Ce${_3}$Os${_4}$Ge$_{13}$ at 2~K is $\approx 0.92 \mu_\mathrm{B}$/Ce. This reduction in the magnetization is attributed to Kondo effect and the crystal electric field effects as observed in other Ce-based compounds \cite{Baumbach2015,Thamizhavel2011,Hegger2000}.  The ac-susceptibility as a function of temperature from 0.1-4.2~K is shown in Fig.~\ref{fig:fig3}(b). The ac-susceptibility rapidly increases below 0.5~K, indicating ferromagnetic ordering of minority Ce$^{3+}$ moments and presence of long range ferromagnetic correlations in Ce${_3}$Os${_4}$Ge$_{13}$.

\begin{figure}[h]
\includegraphics[width=7cm,angle=0]{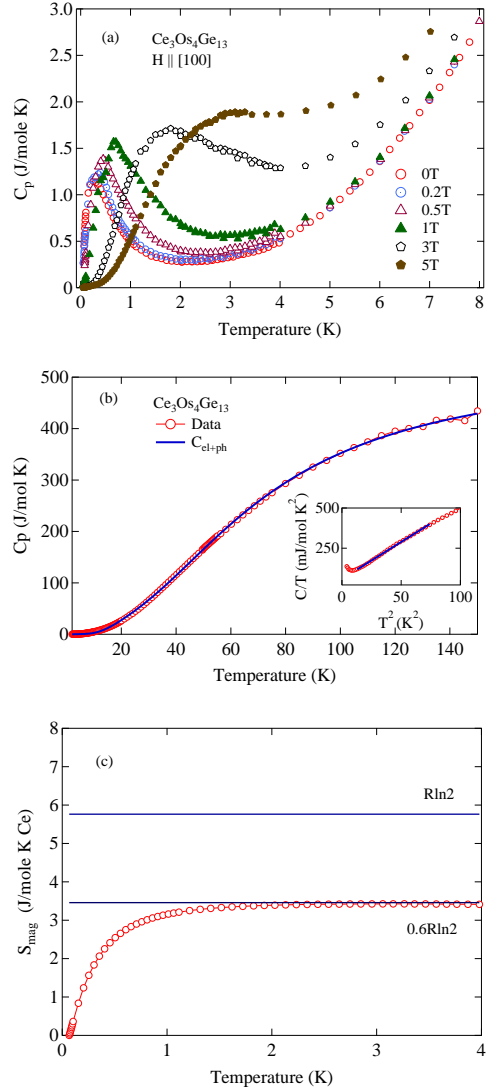}
\caption{(Color online) (a) Low temperature heat capacity $C_{p}(T)$ as a function of temperature ($T$) at different magnetic fields of the Ce${_3}$Os${_4}$Ge$_{13}$ single crystal. With increasing magnetic field ($H$), the peak in $C_{p}(T)$ becomes broader and shifts towards higher temperatures indicating ferromagnetic ordering in Ce${_3}$Os${_4}$Ge$_{13}$. (b) $C_{p}(T)$ ($H=0$~T) as a function of temperature from 0.05-150~K. The solid blue curve is fit to the data using Debye-Einstein model (see text). Solid line in the inset is a linear fit to the $C_{p}(T)$/T vs $T^2$ data at low temperatures. (c) Magnetic entropy ($S_{mag}$) per active Ce moment as a function of temperature.}
\label{fig:fig4}
\end{figure}

This long range magnetic ordering also show up with signature peak in the heat capacity data below 0.5~K. The low temperature heat capacity data at different magnetic fields is presented in Fig.~\ref{fig:fig4}(a). The peak in the heat capacity data clearly reveals the long range magnetic ordering below 0.5~K in zero magnetic field. With the application of magnetic fields, the peak in the heat capacity broadens and shifts to higher temperatures confirming ferromagnetic nature of ordering of Ce$^{3+}$ moments. The height of the heat capacity peak increases with increasing magnetic field and this is not fully understood at the moment. Although, the low temperature heat capacity and its peculiar dependence on magnetic field along with ac-susceptibility data suggest ferromagnetic ordering of minority Ce$^{3+}$ spins below 0.5~K, more sensitive low temperature $\mu$SR and neutron diffraction measurements are required to fully understand the complex nature of magnetic ordering in Ce${_3}$Os${_4}$Ge$_{13}$. The temperature dependence of the heat capacity ($C_{p}$) from 1.8 to 150~K is shown in Fig.~\ref{fig:fig4}(b). 

$C_{p}$ decreases monotonically with lowering the temperature and shows no visible signatures of magnetic orders or structural transitions above 1.8~K. The high temperature heat capacity
is dominated by phonons, which is expected since the magnetic contribution to the heat capacity will be very small (as we have only 0.25 Ce$^{3+}$ moments per formula unit). To quantitatively understand the electronic and phonon contributions to the total heat capacity, we carry out detailed analysis of the heat capacity data using,
\begin{equation}
\label{eqn2}
C_{\rm {el+ph}}(T) = \gamma T +\left[ \alpha C_{\rm Debye}(T) + (1 - \alpha) C_{\rm Einstein}(T)\right],
\end{equation}  
where the first term represents the contribution of the conduction electrons and the the second term represents the phononic contribution to the heat capacity consisting of Debye and Einstein terms. $\alpha$ is a parameter used to determine the relative contributions of Debye and Einstein terms to the phonon heat capacity.

The Debye and Einstein heat capacities are given by the following expressions, 
\begin{equation}
\label{eqn3}
C_{\rm Debye} = 9 n R \left(\frac{T}{\Theta_{\rm D}}\right)^3 \int_0^{\Theta_{\rm D}/T} \frac{x^4 e^x}{(e^x -1)^2} dx,
\end{equation}
and
\begin{equation}
\label{eqn4}
C_{\rm Einstein} = 3 n R \frac{y^2 e^y}{(e^y - 1)^2} ,
\end{equation}

where $\Theta_{\rm D}$ is the Debye temperature, $\Theta_{\rm E}$ is the Einstein temperature, $x = \Theta_{\rm D}/T$ and $y = \Theta_{\rm E}/T$.  The best fit to the $C_{p}$(T) data using Equation~(\ref{eqn2}) reveals that total phonon heat capacity has 88\% contribution from the Debye term and remaining 12\% from the Einstein term. We obtain Debye temperature $\Theta_{\rm D} = 302.7\pm{1.2}$~K and Einstein  temperature $\Theta_{\rm E} = 83.7\pm{1.5}$~K from the fit. $C_{p}/T$ vs $T^2$ data is fitted to the equation $C_{p}/T = \gamma T + \beta T^2$, as presented in the inset of Fig.~\ref{fig:fig4}(b). The value of the normal state Sommerfeld coefficient $\gamma =64.2$~mJ/mol-K$^2$ is obtained from the fit. It is unlikely that the mixed-valent Cerium (Ce$^{4-\delta}$) or the Osmium/Germanium atoms contribute significantly to the electronic heat capacity, therefore the large effective masses are expected to be primarily due to the Ce$^{3+}$ spins at the 2(a) site. Thus, the true effective masses of the trivalent cerium atoms (Ce$^{3+}$) are really four times larger (since a formula unit of Ce${_3}$Os${_4}$Ge$_{13}$ structure consists of 0.25 Ce$^{3+}$ atoms) than the experimentally derived value.\\
The magnetic contribution to the heat capacity was obtained by subtracting the conduction electron and phonon contributions, $C_{mag}=C_p-C_{el+ph}$. The magnetic entropy $S_{mag}$ for Ce$^{3+}$ spins as a function of temperature was obtained by numerically integrating the $C_{mag}/T$ vs T data. The magnetic entropy($S_{mag}$) as a function of temperature is shown in Fig.~\ref{fig:fig4}(c). $S_{mag}$ shows change in slope above $\approx 0.5~K$ and saturates to $0.6R\ln2 = 3.46$ J/mol-K-Ce around ~4~K. This entropy value is a fraction of $R\ln2$, the value for the crystal electric field (CEF) splitted doublet ground state. This suggests doublet ground state of Ce ions produced by CEF splitting of the Hund's rule multiplet. The reduction in entropy of the Ce$^{3+}$ spins is attributed to the Kondo driven hybridization of the Ce$-4f$ and conduction electrons \cite{Baumbach2015, Hegger2000}.\\

\begin{figure}[!h]
\includegraphics[width=7cm,angle=0]{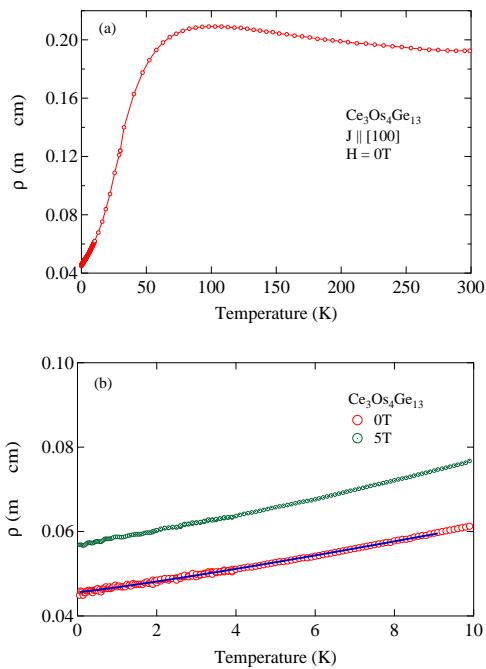}
\caption{(Color online) (a) Temperature dependence of resistivity $\rho(T)$ from 0.05-300~K for $J\parallel [100]$. The resistivity shows a broad peak around 100~K and decreases rapidly at low temperatures. (b) Low temperature resistivity measured in 0 and 5~T magnetic fields. The 0~T data is described by $\rho(T ) = \rho(0) + AT^{1.1}$ showing non-Fermi liquid behavior of the quasiparticles. Comparison of the resistivity values at 0~T and 5~T shows large $125\%$ magnetoresistance at 2~K.}
\label{fig:fig5}
\end{figure}

The temperature dependence of the resistivity ($\rho(T)$) of Ce${_3}$Os${_4}$Ge$_{13}$ single crystal from 0.05-300~K at 0~T and 5~T magnetic fields is presented in Fig.~\ref{fig:fig5}(a). The resistivity increases monotonically with decreasing temperature in the temperature range $300-100~K$ and decreases rapidly below 100~K with the residual resistivity ratio value [$\rho(300)$ K/$\rho(2)$K=4.3]. The behavior of $\rho(T)$ is similar to the one observed in other heavy fermion compounds like CeAl$_3$ \cite{BUSCHOW1970,KAGAYAMA1994}. The broad hump in $\rho(T)$ at higher temperatures is attributed to the Kondo scattering \cite{Kondo1964} of Ce$^{3+}$ moments and conduction electrons. The subsequent fall in $\rho(T)$ is probably due to the development of Kondo coherence \cite{DONIACH1977} below 100~K. A positive magnetoresistance ($\approx 125\%$ in a field of 5~T) is observed at low temperatures as shown in Fig.~\ref{fig:fig5}(b), implying the complex nature of the magnetic correlations in this compound. Positive magnetoresistance is also observed in many other Ce-based Kondo-systems \cite{Coleridge1987, Oomi1992} and is attributed to the formation of the Kondo coherence at low temperatures (T$<<T_K$, where $T_K$ is the Kondo temperature). The low temperature $\rho(T)$ data can be fitted to a power law equation, $\rho(T)=\rho{(0)} + AT^{1.13}$ between 1-10~K, where  $\rho{(0)}$ is the residual resistivity due to static defect in the crystal. We obtain $\rho{(0)}=0.045$~m$\Omega$-cm and $A=1.1\times10^{-3}$~m$\Omega$-cm-K$^{-3}$ from the fit as shown in Fig.~\ref{fig:fig5}(b), suggesting non-Fermi liquid behavior at low temperatures in Ce${_3}$Os${_4}$Ge$_{13}$. From the analysis of the low temperature Curie-Weiss susceptibility and the low temperature resistivity data, we presume that the magnetic correlations are ferromagnetic in nature with long range ordering below 0.5~K. We would like reiterate here, that the magnetism arises due to the minority Ce$^{3+}$ spins at the 2(a) site in the unit cell.

From the above discussion, we establish that Ce${_3}$Os${_4}$Ge$_{13}$ exhibits site disorder driven complex ferromagnetic ordering of minority Ce$^{3+}$ spins (in 2(a) site) below 0.5~K. The existence of two different Ce valence states is a rare occurrence for a metallic system. It is rather unusual for small concentration of Kondo screened Ce spins to undergo bulk magnetic ordering in a site disordered crystal lattice. Furthermore, non-Fermi liquid behavior is observed from the power-law dependence of resistivity and $\log(T)$ dependence of magnetization at low temperatures. The Curie constant determined from the magnetic susceptibility measurements above 140~K and bulk studies indicate that only 8$\%$ of the total Ce ions are in the Ce$^{3+}$ state ordering ferromagnetically below 0.5~K. Low temperature $\mu$SR and Neutron scattering measurements are required to fully understand the complex  magnetic interactions of minority Ce$^{3+}$ spins in this compound. 
\section{Conclusion}
In summary, we have shown complex ferromagnetic ordering of minority spins ($\approx 8\%$ of Ce$^{3+}$) in a site-disordered single crystal of Ce${_3}$Os${_4}$Ge$_{13}$. The structural and magnetization measurements clearly establish that $\approx 8\%$ of Ce atoms are going into the Ge 2(a) site (where Ce is in Ce$^{3+}$ state) from their regular 6(d) site (where Ce is in Ce$^{4-\delta}$ state). The low temperature heat capacity and ac-susceptibility measurements confirm the ferromagnetic ordering of these Ce$^{3+}$ moments below 0.5~K. The heat capacity shows a peak and the real part of the ac-susceptibility shows sharp upturn below 0.5~K. The peak in the heat capacity broadens and shifts towards higher temperature on the application of magnetic field, confirming ferromagnetic nature of the magnetic ordering. The magnetic entropy $S_{mag}$ saturates to $0.6R\log2$ suggesting a crystal field splitted doublet Ce ground state. The low temperature resistivity shows a power-law behavior ($\rho (T)=\rho_0 + AT^{1.13}$). A large 125$\%$ positive magnetoresistance is observed at 2~K in 5~T magnetic field. Low temperature magnetization shows $\log(T)$ behavior ($M(T)=a-b\log(T)$). In addition to the ferromagnetic ordering below 0.5~K, Ce${_3}$Os${_4}$Ge$_{13}$ shows non-Fermi liquid behavior at low temperatures.
\section{Acknowledgments}
We thank Mr. Anil Kumar and Mrs. Ruta Kulkarni for their help during experiments.
\section*{References}
\bibliography{references}
\bibliographystyle{apsrev4-1}
\end{document}